\documentclass{moriond}
\usepackage[compat=1.1.0]{tikz-feynman}
\usepackage{enumitem,color,xcolor}
\usepackage{graphicx,tabularx,tabularray}
\definecolor{dgreen}{rgb}{.0,.6,.0}
\definecolor{lime}{HTML}{A6CE39}
\definecolor{lg}{RGB}{220,220,220} 

\def\be{\begin{equation}}
\def\ee{\end{equation}}
\def\bea{\begin{eqnarray}}
\def\eea{\end{eqnarray}}

\newcommand{\Photo}{\includegraphics[height=35mm]{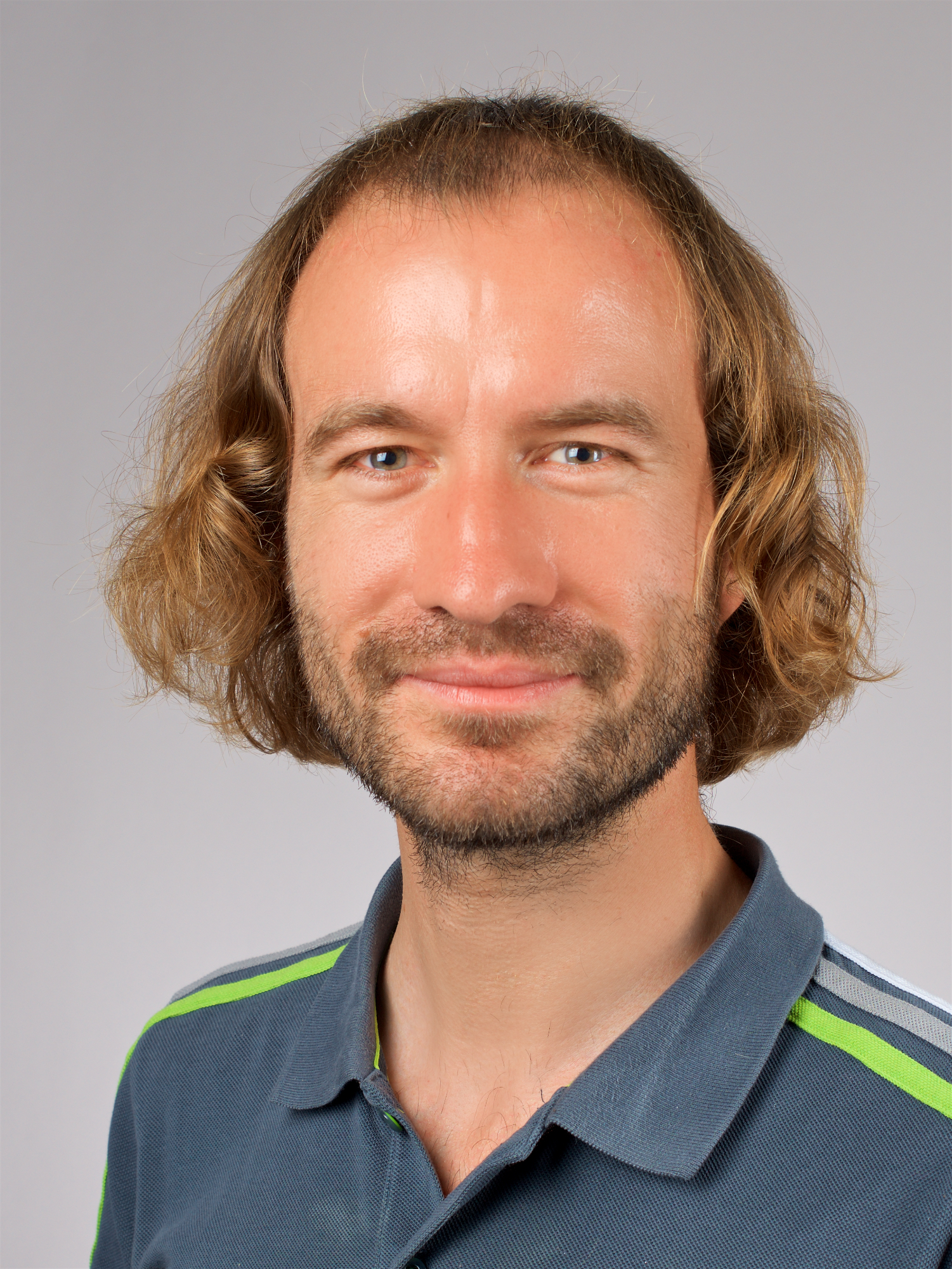}}

\begin{document}
\vspace*{4cm}
\title{New Higgses at the Electroweak Scale and Differential $t\bar t$ Distributions}

\author{Andreas Crivellin}

\address{Physik-Institut, Universität Zürich, Winterthurerstrasse 190, CH–8057 Zürich, Switzerland\\ Paul Scherrer Institut, CH–5232 Villigen PSI, Switzerland}

\maketitle\abstracts{
Indications for new Higgs bosons at 95\,GeV and 152\,GeV with significance of 3.8$\sigma$ and $4.3\sigma$, respectively, have been obtained. While the former contains the inclusive $\gamma\gamma$ channel, the latter is obtained by combining several modes of associated di-photon production, i.e.~$\gamma\gamma+X$ with $X=\ell,\,\ell b,\,{\rm MET},\,\tau,\,...$, within the $\Delta$SM (the Standard Model extended by a real triplet). Such a triplet predicts a positive definite shift in the $W$ mass, in agreement with the current global electroweak fit, and its neutral component decays dominantly to $W$ bosons (for small mixing angles). This offers a connection to $t\bar t$ differential distributions whose experimental signature is $WW b\bar b$: A simplified model with a new scalar $H$, produced via gluon fusion and decaying to $S$ and $S^\prime$ with subsequent (dominant) decays to $WW$ and $b\bar b$, respectively, has the same final state. In fact, adding this new physics polution to the Standard Model describes the leptonic $t\bar t$ distributions better by at least $5.8\sigma$. Furthermore, assuming that $S^\prime$ is an $SU(2)_L$ singlet, the resulting di-photon signal strength is compatible with the 95\,GeV $\gamma\gamma$ measurements. A possible UV completion of this simplified model is the $\Delta$2HDMS, i.e.~a 2HDM supplemented with a singlet and a real triplet, which can successfully accommodate electroweak scale Baryogenesis.}

\section{Introduction}

The Standard Model (SM) cannot be the ultimate theory of physics as, among several shortcomings, it fails to explain the astrophysical observations of Dark Matter and cannot account for the non-vanishing neutrino masses required by neutrino oscillations. Also the minimality of its scalar sector with a single $SU(2)_L$ doublet Higgs giving rise to the mass of all (fundamental) fermions as well as the $W$ and $Z$ bosons is puzzling because it is not guaranteed by any symmetry or consistency requirement. While additional Higgs bosons must play a subleading role in the spontaneous breaking of the electroweak (EW) symmetry due to the constraints from Higgs signal strengths and the global EW fit, the average of the $W$ boson mass measurements is, in fact, slightly higher than the SM prediction ($3.7\sigma$~\cite{deBlas:2022hdk}). Furthermore, many deviations from the SM predictions point towards an extension of the SM scalar sector~\cite{Crivellin:2023zui} and interesting indications for new Higgses with electroweak scale masses of 95\,GeV and 152\,GeV have emerged. This poses the question of whether these hints for narrow resonances are related to the LHC multi-lepton anomalies which appear in Higgs-like typologies in final states with at least two leptons, moderate missing energy and ($b$-)jets~\cite{Fischer:2021sqw}. In particular, differential lepton distributions of top-quark pair production and decay show strong tensions with the SM predictions which might be due to a new physics (NP) contamination involving both the (hypothetical) 95\,GeV and 152\,GeV bosons.

\section{Hints for narrow resonances}

\subsection{95\,GeV ($S^\prime$)}

\begin{figure}[t!]
    \includegraphics[width=1\linewidth]{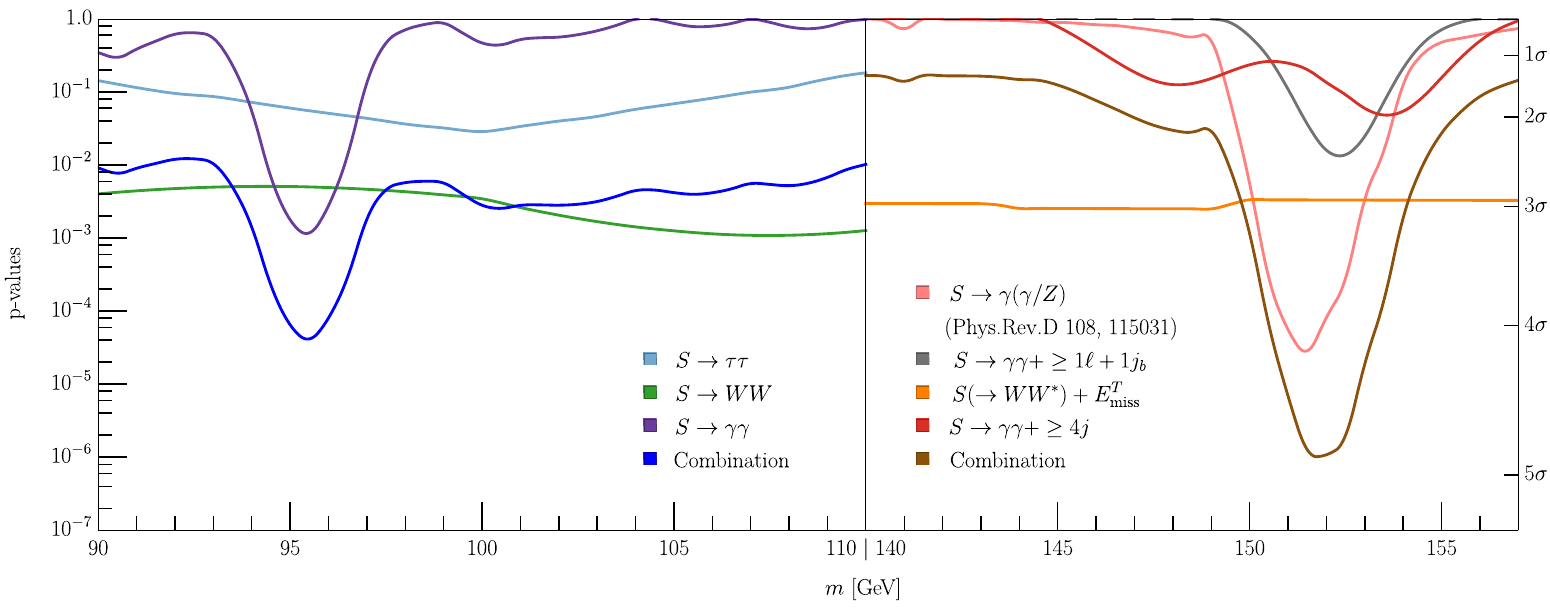}
    \caption{Indications for narrow resonances at the EW scale. While for the low-mass region containing the 95\,GeV excess one relies on inclusive searches, the higher mass region with the 152\,GeV excess has the most significant hints in associated production channels. For the later, the combination is obtained within the simplified model $H\to S S^*$ with $H$ produced via gluon fusion from a top-quark loop and the decay channels $S\to WW,\,b\bar b,\,{\rm MET},\gamma\gamma$ as motivated by the multi-lepton anomalies.}
    \label{combined}
\end{figure}

Combining the CMS~\cite{CMS:2023yay} and ATLAS analyses of $pp\to {S^\prime \to \gamma \gamma}$~\cite{ATLAS:2023jzc} a local significance of $2.9\sigma$ at 95\,GeV is obtained~\cite{Biekotter:2023oen}. While CMS finds a local excess of $3.1\sigma$ in $pp\to S^\prime \to \tau \tau$~\cite{CMS:2022goy}, the side-band of the corresponding SM Higgs ATLAS analysis~\cite{ATLAS:2022yrq} shows no indications of a signal at 95\,GeV, which reduces the CMS significance by a factor $\sqrt{2}$ (assuming that the ATLAS and CMS analyses have similar sensitivity). Recasting the SM Higgs analyses of ${{pp\to S^\prime \to W W^{*}}}$~\cite{Coloretti:2023wng}, a preference of $\approx$2.6$\sigma$ for a non-vanishing cross section at 95\,GeV is found. Finally, LEP reported an excess in $e^+e^-\to Z^*\to(S^\prime\to bb)+Z$ with a local significance of 2.3$\sigma$ at $\approx$98\,GeV~\cite{LEPWorkingGroupforHiggsbosonsearches:2003ing}. Using the latter to reduce the search window to $93\,{\rm GeV}<m_{S^\prime}<103\,$GeV (but not including it directly in the significance calculation) and combining ${S^\prime \to \gamma \gamma}$, ${{S^\prime \to W W^{*}}}$ and $S^\prime \to \tau \tau$ with three degrees of freedom gives a global significance (including the trails-factor) of $3.8\sigma$~\cite{Bhattacharya:2023lmu} (see left side of Fig.~\ref{combined}).

\subsection{152\,GeV ($S$)}

In this mass region, we consider the side bands of SM Higgs analyses. While an $\approx2\sigma$ excess in the inclusive di-photon channel exists at $\approx$152\,GeV~\cite{Crivellin:2021ubm}, the most significant excesses appear in $\gamma\gamma+X$~\cite{ATLAS:2023omk}, i.e.~the associated production of di-photons with $X=\ell,\,{\rm MET},\, \ell b,\,\tau,\,...$. Combining these channels in a simplified model with $pp\to H\to SS^*$ ($m_H\approx 270$\,GeV) and $S\to WW,\,b\bar b,\,{\rm MET},\gamma\gamma$, as motivated by the the multi-lepton anomalies~\cite{vonBuddenbrock:2016rmr,Buddenbrock:2019tua}, a combined significance of $4.7\sigma$ is obtained~\cite{Bhattacharya:2023lmu} (see right plot of Fig.~\ref{combined}). Note that this does not include the $\gamma\gamma+\tau$ channel presented at Moriond EW~\cite{ATLAS-CONF-2024-005} which shows a significant excess around 152\,GeV. 

\begin{figure*}[t!]
\centering
\includegraphics[scale=0.6]{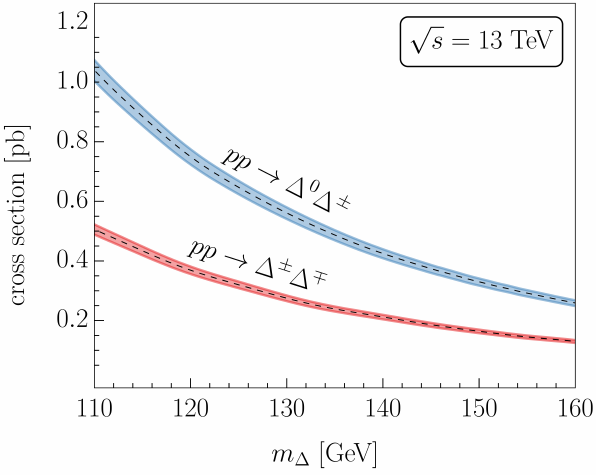}
\hspace{0.3cm}
\includegraphics[scale=0.6]{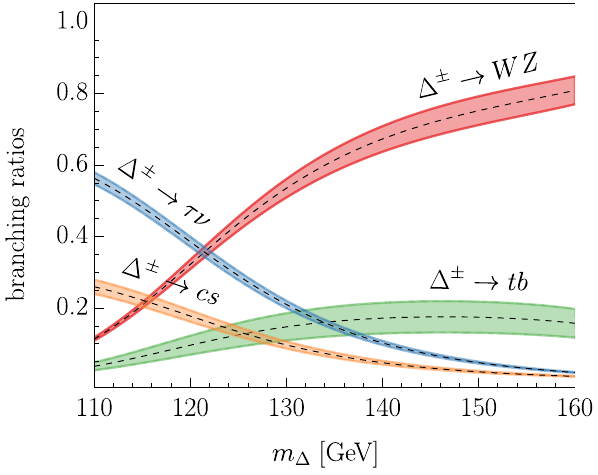}
\caption{Left: Production cross-section for $pp\to \Delta^0 \Delta^\pm$ and $pp\to \Delta^\pm \Delta^\mp$ as a function of the triplet mass. Right: Dominant branching ratios of the charged component $\Delta^\pm$ as a function of its mass. The errors are estimated from the decays for a SM Higgs with a higher (hypothetical) mass from $h\to tt^*,ZZ^*$ and $h\to cc$.}
\label{fig:xsec}
\end{figure*}

Going beyond a simplified model, one can consider the $SU(2)_L$ triplet Higgs with hypercharge 0~\cite{Ross:1975fq} which contains a neutral component $\Delta^0$ and a charged component $\Delta^\pm$ which are approximately degenerate in mass. This model predicts a positive shift in the $W$ mass and leads to $\gamma\gamma+X$ signatures via the Drell-Yan process $pp\to W^*\to \Delta^\pm\Delta^0$~\cite{Ashanujjaman:2024pky} with a cross-section entirely governed by the EW gauge couplings of around half a picobarn (see Fig.~\ref{fig:xsec}, left). Furthermore, since the decay rates of $\Delta^\pm$ depend to a good approximation only on its mass, $WZ$ is dominant for 152\,GeV followed by $tb$ and $\tau\nu$ (see Fig.~\ref{fig:xsec}, right), one can combine all channels with a single degree (the branching ratio of $\Delta^0$ to photons) for a given mass. Including in addition to Ref.~\cite{ATLAS:2023omk} the $\gamma\gamma+\tau$ channel presented at Moriond EW~\cite{ATLAS-CONF-2024-005} a combined significance of $4.3\sigma$ is obtained~\cite{Crivellin:2024uhc} within this simple UV complete model.  

\section{Differential top-quark distributions}

\begin{figure*}[t!]
    \includegraphics[width=0.78\linewidth]{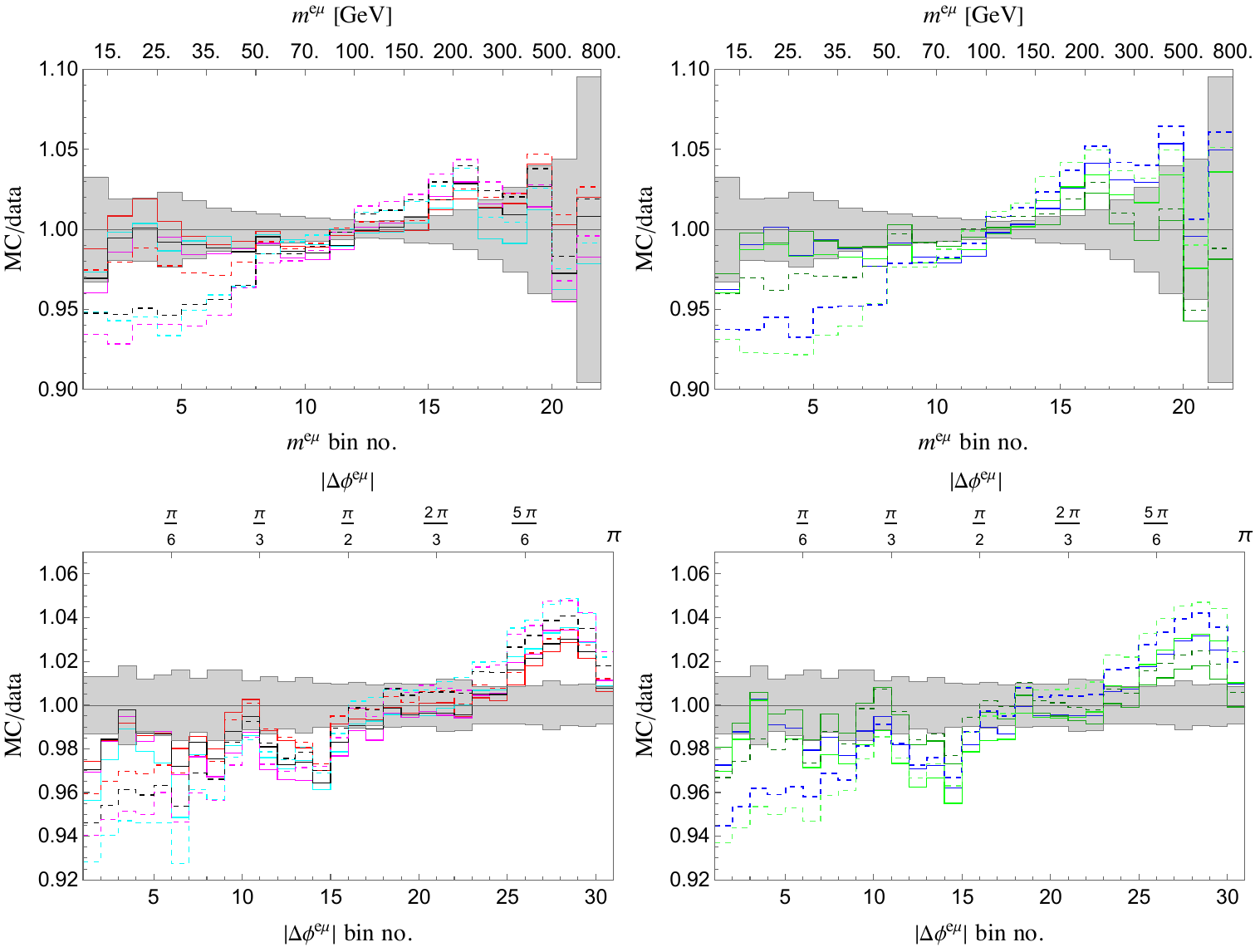}
    \raisebox{0.26\height}{\includegraphics[width=0.21\linewidth]{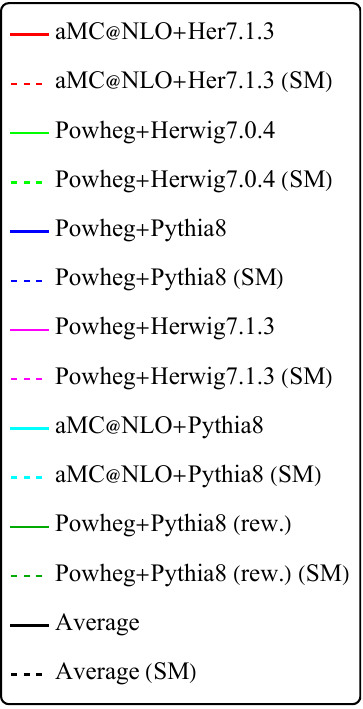}}
    \caption{The dashed coloured lines show the six different SM predictions (MC) normalized to data provided by ATLAS. The solid lines include the NP contribution from our benchmark model obtained by a combined global fit to $m^{e\mu}$ and $\Delta\phi^{e\mu}$ data. The black lines are obtained by averaging the six predictions and the grey band shows the total uncertainty (systematic and statistical). One can see that the agreement between theory and experiment is significantly increased by adding the NP effect.}
    \label{fig:mlldata}
\end{figure*}

The statistically most significant multi-lepton anomaly is encoded in the latest ATLAS analysis of the $t\bar t$ differential cross-sections~\cite{ATLAS:2023gsl}, i.e.~appears in $e^\pm\mu^\mp+b$ final states. The di-lepton mass ($m^{e\mu}$) and the angle between the leptons ($\Delta \phi^{e\mu}$) are the most relevant observables (see Fig.~\ref{fig:mlldata}). Several SM simulations, using different Montecarlo generators and showering, were performed by ATLAS. However, all of them describe data so poorly (see $\chi^2$ values of the SM in Table~\ref{tab:resmass150}) that  Ref.~\cite{ATLAS:2023gsl} concluded: ``No model (SM simulation) can describe all measured distributions within their uncertainties.''

\begin{figure}[t!]
    \centering
    \begin{tikzpicture}[baseline=(current bounding box.center)]
            \begin{feynman}
            \vertex (a);
            \vertex [above left=1.5cm of a] (c) {$g$};
            \vertex [below left=1.5cm of a] (d) {$g$};
            \vertex [right=1.5cm of a] (b) ;
            \vertex [above right=1.5cm of b] (e);
            \vertex [below right=1.5cm of b] (f);
            \vertex [above right=0.75cm of e] (i) {$\bar b$};
            \vertex [below right=0.75cm of e] (j) {$W$};
            \vertex [above right=0.75cm of f] (k) {$W$};
            \vertex [below right=0.75cm of f] (l) {$b$};
            \diagram{
                (d) -- [gluon] (a) -- [gluon] (c);
                (a) -- [gluon, edge label=$g$] (b);
                (f) -- [fermion, edge label=$t$] 
                (b) -- [fermion, edge label=$\bar t$] (e);
                (j) -- [boson] (e) -- [fermion] (i);
                (l) -- [fermion] (f) -- [boson] (k);
            };
        \end{feynman}
            \end{tikzpicture}~~~~~~
                \begin{tikzpicture}[baseline=(current bounding box.center)]
        \begin{feynman}
            \vertex (a);
            \vertex [above left=1.5cm of a] (c) {$g$};
            \vertex [below left=1.5cm of a] (d) {$g$};
            \vertex [right=1.5cm of a] (b) ;
            \vertex [above right=1.5cm of b] (e);
            \vertex [below right=1.5cm of b] (f);
            
            \vertex [above right=0.75cm of e] (i) {$b$};
            \vertex [below right=0.75cm of e] (j) {$\bar{b}$};
            \vertex [above right=0.75cm of f] (k) {$W$};
            \vertex [below right=0.75cm of f] (l) {$W$};
            \diagram{
                (d) -- [gluon] (a) -- [gluon] (c);
                (a) -- [scalar, edge label=$H$] (b);
                (f) -- [scalar, edge label=$S$] 
                (b) -- [scalar, edge label=$S^\prime$] (e);
                (j) -- [fermion] (e) -- [fermion] (i);
                (l) -- [boson] (f) -- [boson] (k);
            };
        \end{feynman}
    \end{tikzpicture}
    \caption{Feynman diagrams showing the leading SM contribution to top pair production and decay as well as the NP process $pp\to H\to SS^\prime$ with $S\to WW$ and $S^\prime\to bb$ contaminating the measurements of $t\bar t$ differential distributions.}
    \label{fig:feynman}
\end{figure}
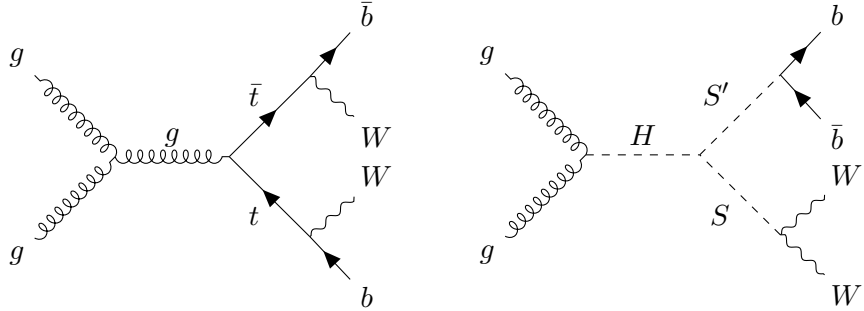

Therefore, one should consider seriously the option that these differential distributions contain a new physics contamination. In fact, the processes $pp\to H\to SS^\prime$ with $S\to WW$ and $S^\prime\to bb$, as shown in Fig.~\ref{fig:feynman} (right), has the same final state as $t\bar t$ production and decay in the SM but with different kinematics. If we fix $m_S=152\,$GeV and $m_{S^\prime}=95\,$GeV from the hints for narrow resonances discussed in the last section, we can fit the cross section for $pp\to H\to SS^\prime\to WWbb$ to data for a given mass of $H$. Since we checked that the impact of varying $m_H$ is small, we chose 270\,GeV as a benchmark point and obtained the results given in table~\ref{tab:resmass150}. From this, one can see that by including the NP effect the agreement with data is significantly improved. Even for the two SM simulations which are in best agreement with the data, the NP hypothesis is preferred over the SM one by at least $5.8\sigma$. 

\begin{table}[!t]
\begin{center}
\begin{tblr}{colspec={Q[c,4.3cm]|Q[c,.6cm]Q[c,.6cm]Q[c,.6cm]Q[c,.6cm]Q[c,2.3cm]}}
Monte Carlo & $\chi _{{\rm{SM}}}^2$ & $\chi _{{\rm{NP}}}^2$ & $\sigma_{{\rm{NP}}}$ & Sig. & $m_S$[GeV] \\
\hline 
\textcolor{blue}{{\rm{Powheg+Pyhtia8}} } 
& 213 & 102 & 9pb & 10.5$\sigma$ & $143-156$\\ 
$\!\!\!$\textcolor{red}{{{\rm{aMC@NLO+Herwig7}}{\rm{.1}}{\rm{.3}}}} 
& 102 & 68 & 5pb & 5.8$\sigma$ & $--$ \\
\textcolor{cyan}{{\rm{aMC@NLO+Pythia8}}} & 291 & 163 & 10pb & 11.3$\sigma$ & 148-157 \\
\textcolor{magenta}{{{\rm{Powheg+Herwig7}}{\rm{.1}}{\rm{.3}}}} & 261 & 126 & 10pb & 11.6$\sigma$ & 149-156\\
\textcolor{dgreen}{{\rm{Powheg+Pythia8~(rew)}}} & 69 & 35 & 5pb & 5.8$\sigma$ & $--$ \\
\textcolor{green}{{{\rm{Powheg+Herwig7}}{\rm{.0}}{\rm{.4}}}} & 294 & 126 & 12pb & 13.0$\sigma$ & 149-156\\
\hline
Average & 182 & 88 & 9pb & 9.6$\sigma$ & 143-157\\
\end{tblr}
\caption{$\chi^2$ values, preferred cross-section ($\sigma_{\rm NP}$) significance (Sig.) etc. for the combined fit to $m^{e\mu}$ and $|\Delta\phi^{e\mu}|$ distributions for the six different SM simulations and their average. The $\chi^2_{\rm NP}$ is for our benchmark scenario with $m_S\approx152\,$GeV while the $m_S$ gives the preferred range of it from the fit, assuming it to be a free parameter. A dash in the $m_S$ column means that the preferred $1\sigma$ rage is wider than $140\,$GeV--$160\,$GeV. The preferred NP cross-section $\sigma_{\rm NP}$ is given for Br$[S^\prime\to b\bar b]\approx100\%$ and Br$[S\to WW]\approx100\%$.}
\label{tab:resmass150}
\end{center}
\end{table}

\begin{figure}[t]
\centering
    \includegraphics[width=0.6\linewidth]{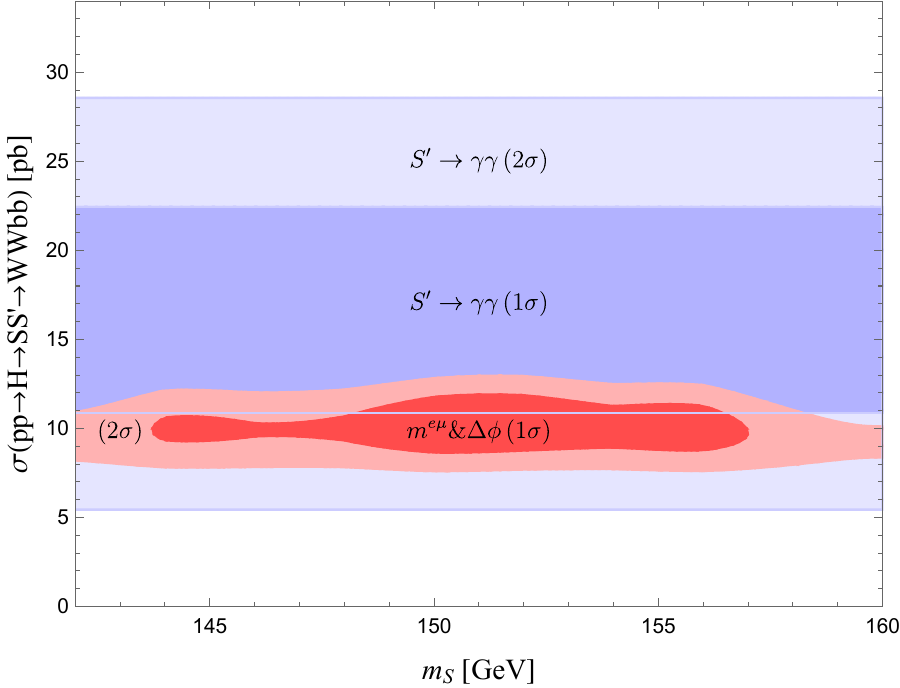}
    \caption{Preferred regions from the $t\bar t$ differential distributions (red) as a function of $m_S$ and the total cross section $pp\to H\to SS^\prime\to WWbb$ assuming $S^\prime$ to be SM-like and Br$[S\to WW]=100\%$. The blue region is preferred by the $95\,$GeV $\gamma\gamma$ signal strength.}
    \label{mSvsEpsNP}
\end{figure}

Allowing now $m_S$, to which the distributions are most sensitive among the new scalar masses, to vary, we show its preferred range in Fig.~\ref{mSvsEpsNP}. We can see that the resulting interval of $\approx(144\,{\rm GeV})$-$\approx(157\,{\rm GeV})$ is compatible with $m_S\approx 152\,$GeV obtained from the $\gamma\gamma+X$ excesses. Furthermore, averaging the six different SM predictions the resulting best fit to $t\bar t $ data is obtained for $\sigma(pp\to H\to SS^\prime \to WW b\bar b)\approx 9$pb. Assuming that $S^\prime$ has SM-like branching ratios (e.g.~is an $SU(2)_L$ singlet), and thus decays to 86\% to $b\bar b$, this results at the same time in a di-photon signal strength in agreement with the 95\,GeV $\gamma\gamma$ excess (see Fig.~\ref{mSvsEpsNP}). Similarly, $S$ should decay dominantly to $WW$ but only suppressed to $ZZ$ (as there is no four-lepton excess at $\approx$152\,GeV), which suggests that it could be the neutral component of an $SU(2)_L$ triplet, further strengthening the connection to the 152\,GeV excess.

\section{Conclusions and Outlook}

Statistically significant indications for new narrow resonances with masses of 95\,GeV and 152\,GeV have been observed in LHC data. While the former appears in inclusive searches, the latter is most pronounced in associated production channels, i.e.~$\gamma\gamma+X$ with $X=\ell,\,{\rm MET},\,\ell b,\,\tau,\,...$. For the 95\,GeV candidate a model-independent significance of 3.8\,$\sigma$ is found, while combining the hints for associated di-photon production within the $\Delta$SM, i.e.~the SM supplemented by a Higgs triplet with $Y=0$, results in a 4.3$\sigma$ excess at 152\,GeV.

The Higgs triplet does not only predict a positive shift in the $W$ mass, as preferred by the current global EW fit, but also provides a possible connection to the multi-lepton anomalies. Since its neutral component decays dominantly to $WW$ (for small mixing angles with the SM Higgs), it could be involved in the explanation of the discrepancies between the SM predictions and measurements of the differential top-quark distributions via the NP contamination $pp\to SS^\prime\to WWbb$. In fact, if one assumes that $S$ is the neutral component of the triplet and $S^\prime$ a singlet, a 95\,GeV di-photon signal strength in agreement with experimental data is obtained.

Recently, we proposed the $\Delta$2HDMS, i.e.~the SM extended by a singlet, a $Y=0$ triplet and a second $Y=1/2$ doublet, as a UV complete model that can account for the top-quark differential distributions and the 95\,GeV and 152\,GeV excesses~\cite{Coloretti:2023yyq}. Interestingly, this model was previously studied in a very different context; it was shown to be above to give rise to EW scale Baryogenesis~\cite{Inoue:2015pza}, i.e.~explain the matter anti-matter asymmetry in the universe. 

\section*{Acknowledgments}

I thank the organizers of Moriond EW and QCD, in particular Gudrun Hiller and Nazila Mahmoudi, for the invitation and the opportunity to present my research. This work is supported by a professorship grant from the Swiss National Science Foundation (No.\ PP00P21\_76884).


\section*{References}

\begin{thebibliography}{99}
\bibitem{deBlas:2022hdk}
J.~de Blas, M.~Pierini, L.~Reina and L.~Silvestrini,
Phys. Rev. Lett. \textbf{129} (2022) no.27, 271801
doi:10.1103/PhysRevLett.129.271801
[arXiv:2204.04204 [hep-ph]].

\bibitem{Fischer:2021sqw}
O.~Fischer, B.~Mellado, S.~Antusch, E.~Bagnaschi, S.~Banerjee, G.~Beck, B.~Belfatto, M.~Bellis, Z.~Berezhiani and M.~Blanke, \textit{et al.}
Eur. Phys. J. C \textbf{82} (2022) no.8, 665
doi:10.1140/epjc/s10052-022-10541-4
[arXiv:2109.06065 [hep-ph]].

\bibitem{Crivellin:2023zui}
A.~Crivellin and B.~Mellado,
Nature Rev. Phys. \textbf{6} (2024) no.5, 294-309
doi:10.1038/s42254-024-00703-6
[arXiv:2309.03870 [hep-ph]].

\bibitem{CMS:2023yay}
 [CMS],
CMS-PAS-HIG-20-002.

\bibitem{ATLAS:2023jzc}
 [ATLAS],
ATLAS-CONF-2023-035.

\bibitem{Biekotter:2023oen}
T.~Biek\"otter, S.~Heinemeyer and G.~Weiglein,
Phys. Rev. D \textbf{109} (2024) no.3, 3
doi:10.1103/PhysRevD.109.035005
[arXiv:2306.03889 [hep-ph]].

\bibitem{CMS:2022goy}
A.~Tumasyan \textit{et al.} [CMS],
JHEP \textbf{07} (2023), 073
doi:10.1007/JHEP07(2023)073
[arXiv:2208.02717 [hep-ex]].

\bibitem{ATLAS:2022yrq}
G.~Aad \textit{et al.} [ATLAS],
JHEP \textbf{08} (2022), 175
doi:10.1007/JHEP08(2022)175
[arXiv:2201.08269 [hep-ex]].

\bibitem{Coloretti:2023wng}
G.~Coloretti, A.~Crivellin, S.~Bhattacharya and B.~Mellado,
Phys. Rev. D \textbf{108} (2023) no.3, 035026
doi:10.1103/PhysRevD.108.035026
[arXiv:2302.07276 [hep-ph]].

\bibitem{LEPWorkingGroupforHiggsbosonsearches:2003ing}
R.~Barate \textit{et al.} [LEP Working Group for Higgs boson searches, ALEPH, DELPHI, L3 and OPAL],
Phys. Lett. B \textbf{565} (2003), 61-75
doi:10.1016/S0370-2693(03)00614-2
[arXiv:hep-ex/0306033 [hep-ex]].

\bibitem{Bhattacharya:2023lmu}
S.~Bhattacharya, G.~Coloretti, A.~Crivellin, S.~E.~Dahbi, Y.~Fang, M.~Kumar and B.~Mellado,
[arXiv:2306.17209 [hep-ph]].

\bibitem{Crivellin:2021ubm}
A.~Crivellin, Y.~Fang, O.~Fischer, S.~Bhattacharya, M.~Kumar, E.~Malwa, B.~Mellado, N.~Rapheeha, X.~Ruan and Q.~Sha,
Phys. Rev. D \textbf{108} (2023) no.11, 115031
doi:10.1103/PhysRevD.108.115031
[arXiv:2109.02650 [hep-ph]].

\bibitem{ATLAS:2023omk}
G.~Aad \textit{et al.} [ATLAS],
JHEP \textbf{07} (2023), 176
doi:10.1007/JHEP07(2023)176
[arXiv:2301.10486 [hep-ex]].

\bibitem{vonBuddenbrock:2016rmr}
S.~von Buddenbrock, N.~Chakrabarty, A.~S.~Cornell, D.~Kar, M.~Kumar, T.~Mandal, B.~Mellado, B.~Mukhopadhyaya, R.~G.~Reed and X.~Ruan,
Eur. Phys. J. C \textbf{76} (2016) no.10, 580
doi:10.1140/epjc/s10052-016-4435-8
[arXiv:1606.01674 [hep-ph]].

\bibitem{Buddenbrock:2019tua}
S.~Buddenbrock, A.~S.~Cornell, Y.~Fang, A.~Fadol Mohammed, M.~Kumar, B.~Mellado and K.~G.~Tomiwa,
JHEP \textbf{10} (2019), 157
doi:10.1007/JHEP10(2019)157
[arXiv:1901.05300 [hep-ph]].

\bibitem{ATLAS-CONF-2024-005}
 [ATLAS],
ATLAS-CONF-2024-005.

\bibitem{Ross:1975fq}
D.~A.~Ross and M.~J.~G.~Veltman,
Nucl. Phys. B \textbf{95} (1975), 135-147
doi:10.1016/0550-3213(75)90485-X

\bibitem{Ashanujjaman:2024pky}
S.~Ashanujjaman, S.~Banik, G.~Coloretti, A.~Crivellin, S.~P.~Maharathy and B.~Mellado,
[arXiv:2402.00101 [hep-ph]].

\bibitem{Crivellin:2024uhc}
A.~Crivellin, S.~Ashanujjaman, S.~Banik, G.~Coloretti, S.~P.~Maharathy and B.~Mellado,
[arXiv:2404.14492 [hep-ph]].

\bibitem{ATLAS:2023gsl}
G.~Aad \textit{et al.} [ATLAS],
JHEP \textbf{07} (2023), 141
doi:10.1007/JHEP07(2023)141
[arXiv:2303.15340 [hep-ex]].

\bibitem{Coloretti:2023yyq}
G.~Coloretti, A.~Crivellin and B.~Mellado,
[arXiv:2312.17314 [hep-ph]].

\bibitem{Inoue:2015pza}
S.~Inoue, G.~Ovanesyan and M.~J.~Ramsey-Musolf,
Phys. Rev. D \textbf{93} (2016), 015013
doi:10.1103/PhysRevD.93.015013
[arXiv:1508.05404 [hep-ph]].

\end{thebibliography}
\end{document}